# TASKS SCHEDULING TECHNIQUE USING LEAGUE CHAMPIONSHIP ALGORITHM FOR MAKESPAN MINIMIZATION IN IAAS CLOUD

Shafi'i Muhammad Abdulhamid*[1]  Muhammad Shafie Abd Latiff[2] and Ismaila Idris[3]

[1, 2, 3] Faculty of Computing, Universiti Teknologi Malaysia, Johor Bahru MALAYSIA.
[1,2] Department of Cyber Security Science, Federal University of Technology Minna, NIGERIA.
E-mail: shafii.abdulhamid@futminna.edu.ng , shafie@utm.my, idris.ismaila95@gmail.com

**ABSTRACT**

Makespan minimization in tasks scheduling of infrastructure as a service (IaaS) cloud is an NP-hard problem. A number of techniques had been used in the past to optimize the makespan time of scheduled tasks in IaaS cloud, which is propotional to the execution cost billed to customers. In this paper, we proposed a League Championship Algorithm (LCA) based makespan time minimization scheduling technique in IaaS cloud. The LCA is a sports-inspired population based algorithmic framework for global optimization over a continuous search space. Three other existing algorithms that is, First Come First Served (FCFS), Last Job First (LJF) and Best Effort First (BEF) were used to evaluate the performance of the proposed algorithm. All algorithms under consideration assumed to be non-preemptive. The results obtained shows that, the LCA scheduling technique perform moderately better than the other algorithms in minimizing the makespan time of scheduled tasks in IaaS cloud.

**INTRODUCTION**

In 2009, Kashan [1] introduced a novel computational intelligence algorithm called the league championship algorithm (LCA). It is a novel optimization technique designed based on the inspiration of soccer competitions in a championship league. The new algorithm is a population based algorithmic scheme for global optimization over a continuous search space. The LCA was designed to be a stochastic population based algorithm for continuous global optimization which tries to mimic a championship situation where synthetic football clubs participate in an artificial league for a period of time. The LCA algorithm has been used in many areas and performed creditably well as compared to other known optimization schemes or heuristics algorithms [2,3].

The cloud computing services can be largely divided into three; the software as a service (SaaS), the platform as a service (PaaS) and the infrastructure as a service (IaaS). The IaaS cloud provides computational resources such as the virtual machines (VMs) to cloud users on demand [4,5]. Tasks scheduling optimization had been an area of research in IaaS cloud because it is an NP-hard problem. Nevertheless, the autonomous attribute and the resource heterogeneity within the clouds and the VM execution necessitate different schemes for task scheduling in the IaaS cloud computing to be used and tested in order to minimize the makespan time. The makespan time is directly responsible for the tasks execution cost in this environment [6,7].

The main objective of this research paper is to propose a tasks scheduling technique in IaaS cloud computing using the LCA, to minimize the makespan. Section II reviewes some related literatures in LCA and also in tasks scheduling in IaaS cloud. Section III puts describes the materials and proposed method from tasks scheduling in environment by enhancing the LCA add-on of winner/looser determination. Section IV presents the experiment setup and results and section V presents the conclusion and future works.

**RELATED LITERATURES**

Stephen and PVGD [8] presented a proposed method that provides some theoretical frame work to use multiple optimization techniques at the same time in a distinct optimization problem. This new method had been expperimented to solve image enhancement problem in a fingerprint. But before then, it was Kashan [9] that first introduced a new evolutionary algorithm known as League Championship Algorithm (LCA) for global optimization, which imitates the sport league championships. It is a new algorithm for numerical function optimization. Kashan and Karimi [9] tests the effectiveness of the proposed optimization algorithm by measuring the test functions from a recognized yardstick, usually adopted to authenticate new constraint-handling algorithms strategy. Sebastián and Isabel [10] presents an implementation of the LCA in a Job Shop scheduling in an industrial situation. Sun, Wang, Li, Wu, Huang and Wang [11] also proposed a market surplus and overall reputation as the optimization objectives and use LCA to decide the winner in an auction, which realizes dynamic, efficient and combinatorial allocation of resources in the cloud environment.

Jacob, Jeyakrishanan and Sengottuvelan [12] presented a bacterial foraging optimization scheme utilized for the scheduling of the resources in the cloud computing system. The outcome of the evaluation shows that the proposed scheme can minimize the usage price, makespan and also maximize the reliability of the scheduling scheme. Diangang, Zhiya, Lin and Liu [13] developed a VM scheduling algorithm in IaaS cloud environment. Shen, Deng, Iosup and Epema [14] puts forward a classification of cloud-based, online, crossbreed scheduling techniques that minimizes usage price by using both on-demand and reserved case in point.

Salot [15] surveyed different scheduling techniques and problems related to them in cloud system. The paper shows that tasks scheduling are very vital in the cloud system because clients have to pay for resources used measured by the period of access. Therefore, proficient usage of cloud resources must be important and for that scheduling plays an important function to get optimum gains from the resources. Sun, Ji, Yue and Yang [16] developed a VM scheduling technique and disaster recovery algorithm for IaaS cloud environment using a runtime and average usage of the three layers of IaaS cloud. Pawar and Kapgate [17] reviewed different schemes and algorithms available for VM scheduling. The review gave an insight into the uniqueness of VM to resolve built up load in an efficient scheduling and management. Zhan and Huo [18] projected an improve particle swarm optimization scheme for resources scheduling technique in cloud computing environment. The outcome of the evaluation shows that the proposed technique can minimize the job average execution time, and increase the rate of availability of resources in the environment.

## MATERIALS AND METHOD

In order to achive the level of optimization needed, the algorithm need to be enhanced. The LCA based task scheduling technique was developed by modifying the original LCA metaheuristic algorithm inspired based on the metaphor of sports championship in a round-robin sport leagues. The LCA implementation steps are guided by the six idealized rules [1, 19];

1) Idealized rule 1: It is more probable that a team with better playing strength triumphs in a match. The phrase "playing strength" means the ability of one team to defeat another team.
2) Idealized rule 2: The result of a match is not predictable given known the teams' playing strength entirely.
3) Idealized rule 3: The chance that team $i$ defeats team $j$ is assumed to be the same from both teams point of view.
4) Idealized rule 4: The result of the match is only win or loss. A tie is not considered in the basic version of LCA (We will later break this assumption via inclusion of the tie outcome, when introducing other variants of the algorithm).
5) Idealized rule 5: If team $i$ defeats team $j$, any strong point helped team $i$ to win has a double weakness caused team $j$ to lose. This is, any weakness is a deficient in a particular strength. An implicit implication of this rule is that while the match outcome is imputed to chance, teams may not believe it technically.
6) Idealized rule 6: Teams only concentrate on their immediate next match without regards tothe other future schedules. Formation settings are done just based on the previous week events.

In order to achive optimization with the proposed algorithm (LCA) in scheduling the cloud tasks, we must first have to match the corresponding variables or parameters of the two systems. To achive this, a simple comparison was made with the variables of a known evolutionary algorithm (EA) and the following matching was achived;

**Table-1.** Parameters Matching

| LCA | EA |
|---|---|
| League $L$ | population |
| week $t$ | Iteration |
| team $i$ | $i^{th}$ member in the population |
| formation $X_i^t$ | solution |
| playing strength $f(X_i^t)$ | fitness value |
| number of seasons S | maximum iterations |

### The Winner/Loser Determination

The LCA has a number of enhancement add-ons and one important add-on is the winner/loser determination feature. In this paper, we used this add-on in determining which task is scheduled on which VM in the IaaS cloud. In a normal league system, teams play each other weekly and their game result is evaluated on the basis of win/loss/tie for each of the teams. For instance, in football league, each club is to get three points for win, zero for l oss and one for draw/tie. By ignoring, the irregular abnormalities which may ensure even outstanding clubs in a variety of unsuccessful outcomes, it is probable that a more dominant club having a superior playing pattern defeats the lesser team. In an ideal league situation that is free from uncertainty effects, an assumption can be easily made for a linear correlation between the playing pattern of a club and the result of its matches. Utilizing the playing power condition, the winner/loser in LCA is determined in a stochastic approach with criteria that the probability of winning for a club is relative to its degree of fit. Given teams $i$ and $j$ playing a league match at week $t$, with their formations $X_i^t$ and $X_j^t$ and playing powers $f(X_i^t)$ and $f(X_j^t)$, correspondingly. Let $P_i^t$ represents the probability of team $i$ to defeat team $j$ at week $t$ ($P_i^t$ is defined respectively). Given also $\hat{f}$ be an ideal value (e.g., a lower limit on the best value).

$$\frac{f(x_i^t) - \hat{f}}{f(x_j^t) - \hat{f}} = \frac{p_i^t}{p_j^t} \qquad (1)$$

From the idealized 3 rule we can also write:

$$p_i^t + p_j^t = 1 \qquad (2)$$

From equations (1) and (2) above we solve for $P_i^t$

$$p_i^t = \frac{f(x_j^t) - \hat{f}}{f(x_j^t) + f(x_i^t) - 2\hat{f}} \qquad (3)$$

In order to find the winner or loser, a random number in [0,1] is generated; if the generated number is less than or equal to $P_i^t$, team $i$ wins and team $j$ loses; otherwise $j$ wins and $i$ loses. This method of finding the win or lose is in line with the idealized rules. If $f(X_i^t)$ be arbitrarily closed to $f(X_j^t)$, then $P_i^t$ can be arbitrarily closed to 1/2. Moreover, if $f(X_j^t)$ becomes far greater than $f(X_i^t)$, namely $f(X_j^t) \gg f(X_i^t)$, then $P_i^t$ approaches to one. Then, the value of $\hat{f}$ may be unavailable in the feature, we use from the best function value found so far (i.e., $\hat{f}^t = \min_{i=1,\dots,L}\{f(B_i^t)\}$. Figure-1 shows the LCA flowchart.

the first task execution to the end of the last task execution in the schedule. It assumes that the tasks are ready at time zero and resources are continuously available during the whole scheduling. Mathematically, makespan can be expressed as;

$$Makespan\ C_{max} = \max\{C_i^{'}\} = \max\{C_1^{'}, C_2^{'}, ..., C_n^{'}\} \quad (4)$$

where, $C_i^{'}$ is the completion of task *i*. The lesser the makespan the better the efficiency of the algorithm, meaning less time is taken to execute the algorithm.

**EXPERIMENT AND RESULTS**

The parameter used for measuring the scheduling algorithms in this experiment are based on the makespan time of the tasks execution. The aim of the experiment is to see how best we can go in minimizing the makespan time of all the schemes under consideration. Three other existing scheduers i.e. First Come First Served (FCFS), Last Job First (LJF) and Best Effort First (BEF) were used to compare and evaluate the performance of the proposed algorithm based on their makespan time. The dataset was formed by using the Delft University of Technology workload traces 200-500 million instructions in MATLAB. The experiment was performed by varying the number of tasks sent in the IaaS cloud from 20 to 180. All four algorithms assumed to be non-preemptive. Figure 2 shows a prototype of the experimental setup (see Figure-2).

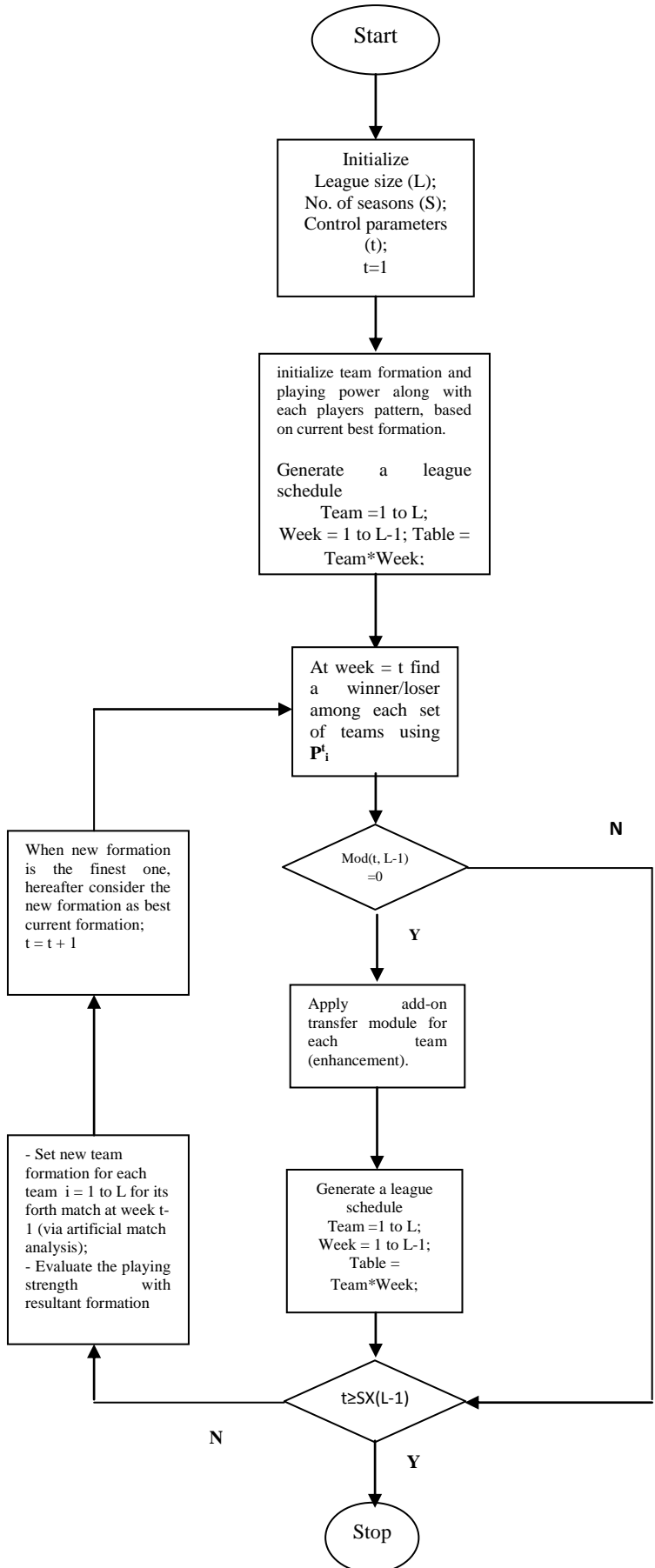

**Figure -1**. Flowchart of the LCA

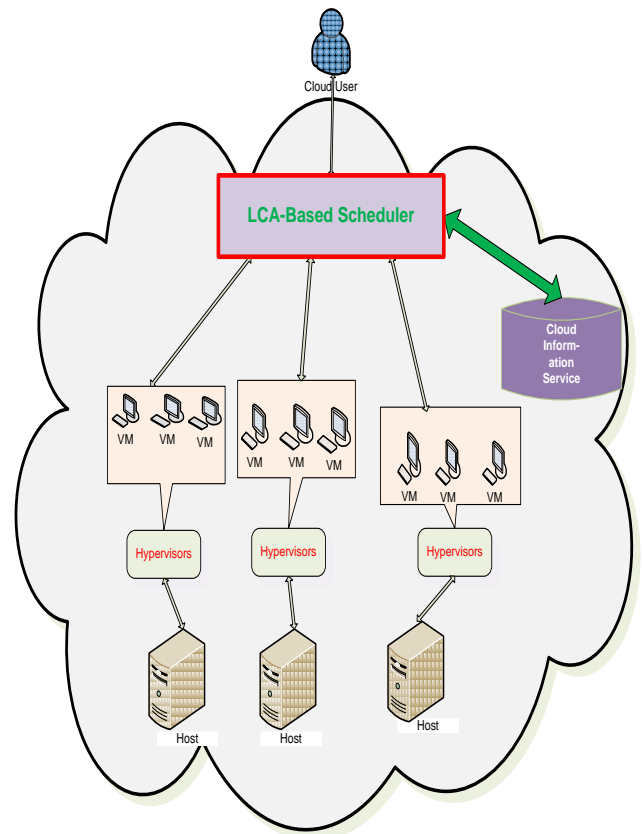

**Figure- 2**. LCA - Task Scheduler in IaaS Cloud

**Performance Metrics**

The makespan time is the maximum completion time of a task. It is also described as the peroid from the start of

**Table -2.** Experimental Parameters

| Parameters | Experimental Values |
|---|---|
| No. of tasks | 20 to 180 |
| No. of VM nodes | 20 |
| No. of Schedulers used | 4 |
| Task workload | 200-500 million instructions |

The experiment was repeated nine times and the average total makespan time for each of the algorithms was captured and tabulated. The makespan time was also computed at different intervals of the experiment. Figure-3 shows the graph of makespan at different tasks execution interval and also with different schedulers.

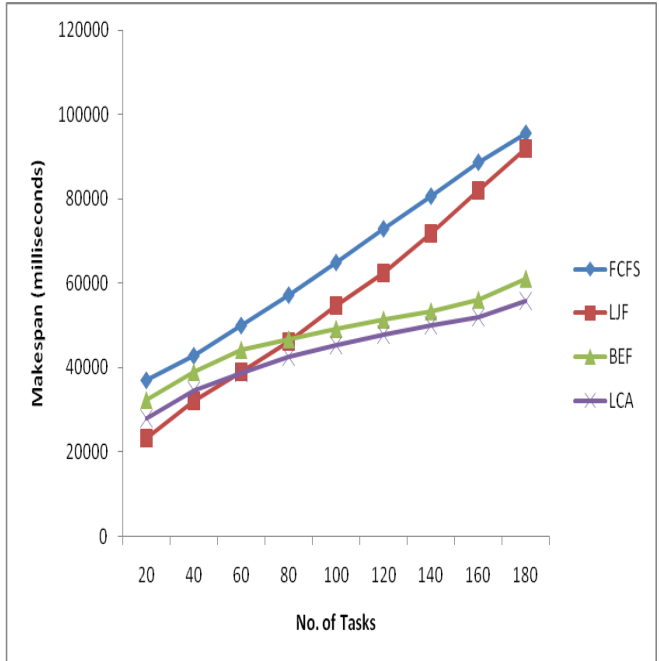

**Figure- 3.** Makespan of Different Schedulers on Different Tasks Interval

Figure- 3 shows the makespan times as calculated by the four scheduling schemes. The makespan time as processed by the LCA scheduling algorithm is lesser than the other three algorithms, i.e. FCFS, LJF and BEF, especially as the number of tasks increases. The FCFS has the highest makespan time amongst the algorithms under consideration. This results obtained from the IaaS cloud environment also shows that, LCA scheduling algorithm perform moderately better than the FCFS and the BEF algorithms throughout the experiment, but only outperformed the LJF as the number of tasks increases. The implication of this result is that, the proposed LCA scheduling scheme will help the cloud customers to save more money while using the cloud. This is because the algorithm helps to reduce the makespan time which is the maximum completion time of tasks, making the customers to spend lesser time in the pay per use IaaS cloud.

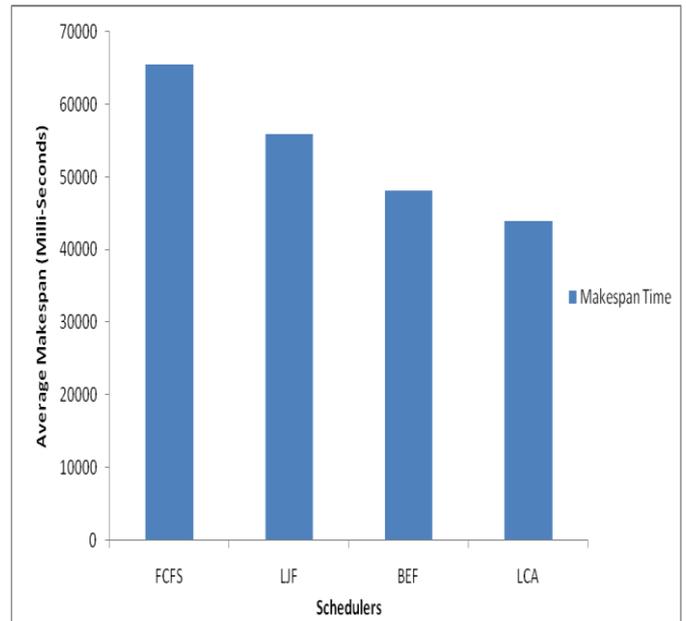

**Figure-4**. Average Makespan Time for Schedulers

Figure 4 shows the average makespan time for all the four algorithms as computed. The FCFS, LJF and BEF scheduling algorithms presented a very high average makespan times as compared to LCA. This result implies that the cloud customers will be paying less for accessing the cloud when using the proposed LCA scheduling technique. Since the makespan time is directly propotional to the amount paid for service.

**CONCLUSION AND FUTURE WORKS**

Minimizing the makespan of tasks scheduling in IaaS cloud is an NP-hard problem that had been attempted with different techniques in the past, but this is the first time the LCA was used to solve this issue. We started with a brief survey of the previous proposed techniques, then put forward a LCA-based makespan minimization scheduling scheme. The experiment shows that, the LCA method produced a lesser makespan than the FCFS, LJF and BEF. This implies a great prospects of performing well in this area as it had performed in solving other NP-complete problems in other areas of research. From this results, it shows that, the LCA helps cloud customers to saves cost for the time used than the LJF, BEF or the FCFS scheduling algorithms, as it take lesser time for the makespan to process the tasks. The LCA is a new sport-based optimization technique that has the potential of adaptation in various fields of research.

We also wish to suggest further reseaches to be carriedout to extend this proposed algorithm into other areas of research such as search techniques in big data, cognitive engeeniring design problem, chaotic sequences in some engineering design, routing problem in distributed networks, learning the Structure of Bayesian networks, assignment problem in graph coloring and other known NP-hard problems.

**ACKNOWLEDGMENTS**

The first author would like to express his appreciation for the support of Universiti Teknologi Malaysia (UTM) International Doctoral Fellowship (IDF), while the second author would like to express his gratitude for the UTM, Research University Grand Q.J130000.2528.05H87 sponsorship for this research. We also wish to appreaciate the Delft University of Technology for providing the workload archive used in this research and Dr. Kashan A. H. for his attention and assistance on research materials.